\documentclass[journal=nalefd,manuscript=article]{achemso}

\usepackage{chemformula} 
\usepackage[T1]{fontenc} 
\usepackage{bm}
\usepackage[normalem]{ulem}
\usepackage[T1]{fontenc}
\usepackage[latin9]{inputenc}
\usepackage{color}
\usepackage{units}
\usepackage{amssymb}
\usepackage{graphicx}
\usepackage{esint}
\usepackage{bm}
\usepackage{natbib}
\usepackage{ulem}
\usepackage{babel}
\usepackage{soul}
\usepackage[colorlinks=true, citecolor=red]{hyperref}

\title{Symmetry broken states at high displacement fields in ABA trilayer graphene}

\author{Simrandeep Kaur$^1$, Unmesh Ghorai$^2$, Abhisek Samanta$^3$, Kenji Watanabe$^4$, Takashi Taniguchi$^5$,  Rajdeep Sensarma$^6$, Aveek Bid$^1$}
	\email{aveek@iisc.ac.in}
\affiliation{$^{1}$Department of Physics, Indian Institute of Science, Bangalore 560012, India\\
$^2$ School of Physics and Astronomy, Tel Aviv University, Tel Aviv 6997801, Israel\\
$^3$ Department of Physics, Indian Institute of Technology Gandhinagar, Gujarat 382355, India\\
$^4$ Research Center for Functional Materials, National Institute for Materials Science, 1-1 Namiki, Tsukuba 305-0044, Japan \\
$^5$ International Center for Materials Nanoarchitectonics, National Institute for Materials Science, 1-1 Namiki, Tsukuba 305-0044, Japan\\
$^{6}$ Department of Theoretical Physics, Tata Institute of Fundamental Research, Homi Bhabha Road, Mumbai, 400005, India}

\begin{document}
\begin{abstract}
In this Letter, we present a comprehensive study of magnetotransport in high-mobility trilayer graphene (TLG) devices under a transverse displacement field, focusing on symmetry-broken Landau levels (LLs) from monolayer-like and bilayer-like bands. A striking displacement-field-induced enhancement of the Land\'e g-factor is observed in the zeroth Landau level of the monolayer-like band, highlighting the role of strong electron-electron interactions. Additionally, we find a rich landscape of LL crossings in the Dirac gully region, accompanied by phase transitions between spin-, gully-, and valley-polarized LLs. These experimental observations are successfully modeled using calculations based on optimized tight-binding parameters. Furthermore, our results reveal significant particle-hole asymmetry in the sequence of LLs in the Dirac gullies, attributed to differing g-factor values for electrons and holes. This asymmetry underscores the limitations of non-interacting models in capturing the complexities of strongly correlated multiband systems. This work provides new insights into the interplay of symmetry-breaking mechanisms and strong correlations in Bernal-stacked trilayer graphene, advancing our understanding of quantum transport phenomena in multiband systems.
\end{abstract}
\maketitle
	

\section{Introduction}
Bernal-stacked trilayer graphene (TLG) is a distinctive condensed matter platform to investigate lattice symmetries and tune them using external parameters like magnetic and displacement fields~\cite{PhysRevLett.117.066601, PhysRevLett.121.056801,datta2017strong, PhysRevLett.117.076807,doi:10.1073/pnas.1820835116}. In its pristine form, TLG is a semi-metallic multiband system formed by a massless monolayer-like (ML-like) band and massive bilayer-like (BL-like) bands (Fig.~\ref{fig:fig1}(a)). These bands are protected from inter-mixing by the lattice mirror symmetry~\cite{PhysRevB.87.115422, PhysRevB.79.125443, PhysRevB.101.245411}. An external perpendicular displacement field $D$ introduces a potential difference $\Delta_1$ between its top and bottom carbon layers, breaking the mirror symmetry and effectively hybridizing the ML-like and the BL-like bands~\cite{ PhysRevB.87.115422, PhysRevLett.121.056801}. In the single-particle picture, $\Delta_1$, along with the interlayer skewed hopping term $\gamma_3$, cause strong trigonal warping of the low-energy spectrum, producing a set of additional off-center low-energy Dirac cones with $C_3$ rotational symmetry at the corners of the Brillouin zone~\cite{PhysRevLett.121.167601, PhysRevB.87.115422, winterer2022spontaneous, PhysRevB.101.245411}. These additional Dirac cones are termed Dirac Gullies and are labeled $\mathrm{T_1}$ and $\mathrm{T_2}$ over the electron side, and $\mathrm{T_3}$ and $\mathrm{T_4}$ over the hole side (Fig.~\ref{fig:fig1}(b)). 

An external electric field can tune the strength of these Dirac Gullies in the momentum space. Notable, the electric field also modulates their effective mass and can induce gap closures, significantly changing their structure at low energies \cite{PhysRevB.87.115422, PhysRevB.101.245411}. Beyond these single-particle effects, interactions can spontaneously break the $C_3$ rotational symmetry. This leads to a Gully polarized ground state phase with the charge carriers occupying a specific off-center Dirac cone~\cite{PhysRevLett.121.167601, PhysRevB.101.245411, winterer2022spontaneous}. Furthermore, with increasing perpendicular magnetic field $B$, inter-gully tunneling becomes prominent, lifting the triplet degeneracy. These emergent Dirac Gullies have been predicted to host chiral edge states \cite{PhysRevB.87.085424}; these were experimentally confirmed recently \cite{PhysRevLett.132.096301}. 

The low-energy physics of TLG is thus replete with $B$ and $D$-induced transitions between phases characterized by different valleys, electronic spin, and Gully degrees of freedom. These transitions are expected to be strongly affected by the electronic interactions in the system, leading to a non-trivial Landau-level spectrum.  The resultant complex evolution of the Fermi surface topology across these transitions can presumably enable a quantitative assessment of electron-electron interactions in this system. Despite robust predictions, there are few experimental studies of this aspect of these transitions~\cite{PhysRevLett.121.167601, winterer2022spontaneous}. This is partly due to the need for devices with exceptionally low disorder where the subtle differences between the closely lying ground states may be discerned. 

This Letter reports detailed magnetotransport studies on very high-mobility TLG devices. We explore the effect of a transverse displacement field on symmetry-broken Landau levels (LLs) originating from the monolayer-like and the bilayer-like bands. We observe a surprising displacement-field-induced enhancement of the Land\'e g-factor in the zeroth Landau level of the monolayer-like band. This enhancement is accompanied by a spontaneous SU(4) symmetry breaking in the bilayer-like LLs, which we attribute to strong electron-electron interactions. Furthermore, we investigate the rich pattern of LL crossings in the Dirac Gully region and the resultant multiple-phase transitions between spin, Gully, and valley-polarized LLs. We successfully map all the experimentally observed transport features with our calculations based on optimized tight-binding parameters~\cite{zibrov2018emergent, PhysRevB.101.245411}. Interestingly, our experimental results reveal a pronounced asymmetry in the evolution of LLs in the Dirac gullies at positive and negative energies;  we establish that this particle-hole asymmetry arises from the stark distinction in the values of g-factor for electrons and holes. Our study highlights the inadequacy of a single-particle picture in studying this strongly correlated system.

\section{Results}

\begin{figure*}[t]
	\includegraphics[width=\columnwidth]{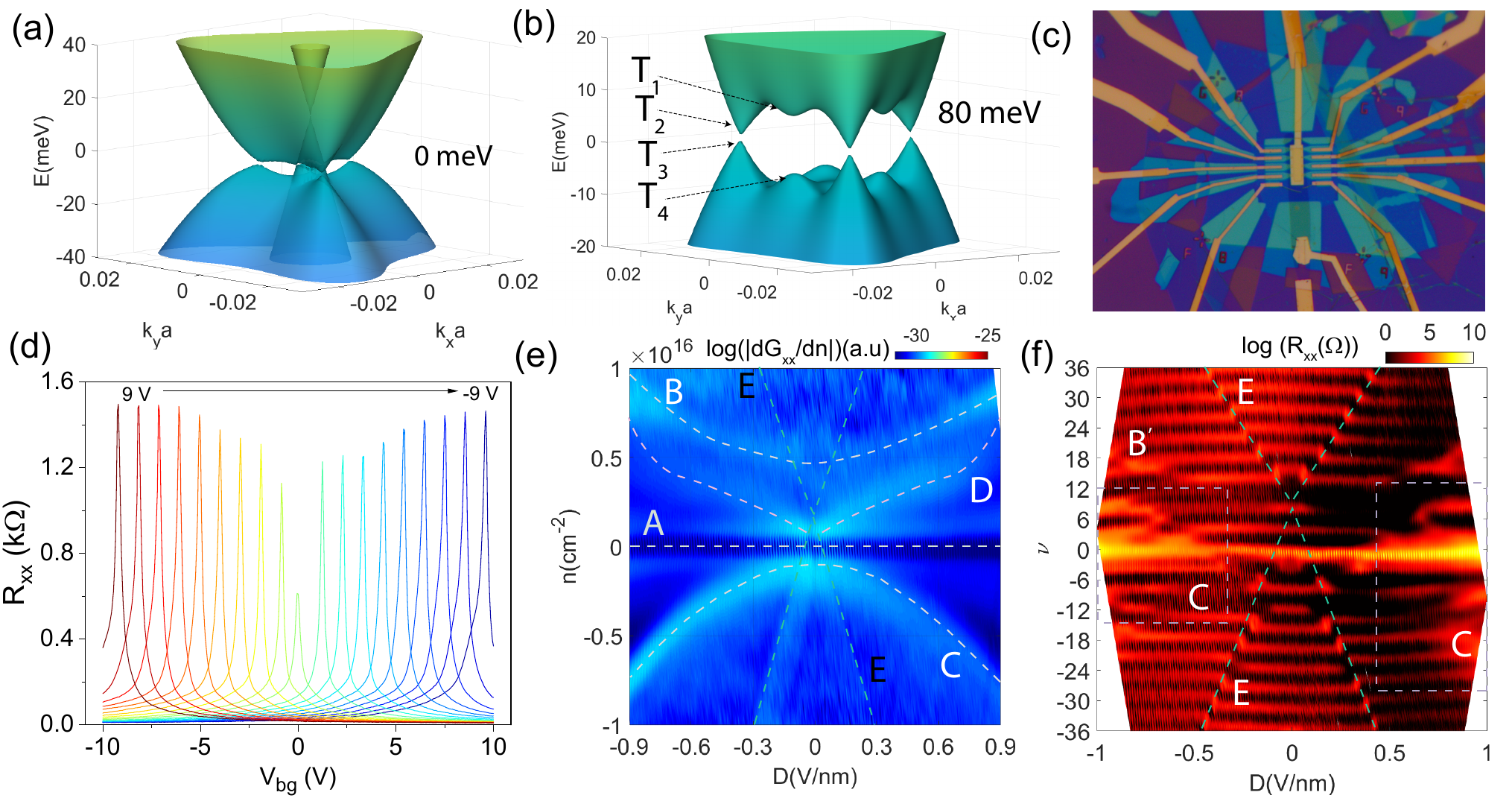}
	\caption{\textbf{Device characteristics:} (a) Calculated ABA trilayer graphene band structure at $D =0$ meV. (b) Calculated band structure of ABA trilayer graphene at $D =80$ meV. $\mathrm{T_1}$ and  $\mathrm{T_3}$ Gullies are hosted by $K$ valley, while $\mathrm{T_2}$ and $\mathrm{T_4}$ Gullies are in the $K^\prime$ valley~\cite{PhysRevB.101.245411}. (c) Optical image of the ABA trilayer graphene device. (d) Plots of $R_{xx}$ versus $V_{bg}$ for representative values $V_{tg}$ at $T=2$ K and $B=0$ T. (e) Contour map of the derivative of the conductance, d$G_{xx}/$d$n$ versus $n$ and $D$. The dotted lines show the evolution of the charge neutrality point (\textbf{A}), the extrema of the monolayer-like bands (\textbf{E}), and several Lifshitz transitions (\textbf{B}, \textbf{C} and \textbf{D}).  (f) Contour map of $R_{xx}$ at $B=2$ T. }
	\label{fig:fig1}
\end{figure*}

Bernal-stacked trilayer graphene flakes (hereafter referred to as TLG), mechanically exfoliated on \ch{SiO2}, were identified through optical contrast and Raman spectroscopy~\cite{Cong2011}. TLG devices encapsulated with hBN with graphite gates (Fig.~\ref{fig:fig1}(c)) were prepared using the dry transfer technique~\cite{Pizzocchero2016,doi:10.1126/science.1244358, Kaur2024}. Dual electrostatic gates were used to tune the number density $n$
and the displacement field $D$ simultaneously~(Supplementary Information, section \textbf{S1}). Line plots of longitudinal resistance versus back gate voltage at different top gate voltages shown in Fig.~\ref{fig:fig1}(d) confirm the device channel to be of ABA  stacked TLG, evidenced by the minimal change in the resistance with displacement fields \cite{Zou2013}.

A contour map of the derivative of the longitudinal conductance d$G_{xx}/$d$n (B=0)$ at $T=20$ mK in the charge carrier density $n - D$ plane is shown in Fig.~\ref{fig:fig1}(e).  We encounter various discontinuities in the d$G_{xx}/$d$n$ that evolve with the displacement field and the number density. The charge neutrality point is the dashed line marked \textbf{A}. With increasing $D$, the monolayer-like bands are predicted to shift to higher energies \cite{Zhou2023, PhysRevLett.121.056801}. The two linearly dispersing lines on both the electron and hole sides (marked \textbf{E}) trace this evolution of the extrema of the monolayer-like bands with $D$. The features marked by lines \textbf{B}, \textbf{C}, and \textbf{D} arise from the multiple Lifshitz transitions due to modification in the BL-like bands~\cite{PhysRevLett.121.167601}. Although seen through Quantum capacitance measurements, these signatures of the Lifshitz transition in TLG had eluded previous transport studies. The detailed Fermi contours in different regions of the $n-D$ plot are shown in the Supplementary Information section \textbf{S2}. Data obtained on another device with qualitatively similar observations are shown in Supplementary Information section \textbf{S3}.

In the Quantum Hall regime, the Lifshitz transitions lead to changes in the LL filling sequence and are discernible as sharp discontinuities in the Landau fan diagrams. Fig.~\ref{fig:fig1}(f) shows the evolution of the minima of $R_{xx}$ in a perpendicular magnetic field $B=2$~T in the $\nu$--$D$ plane ($\nu= nh/eB$ is the LL filling factor, $h$: Planck's constant and $e$: magnitude of electronic charge). The sequence of $\nu$ changes abruptly by $2$ across the dashed line \textbf{E} as a consequence of the LL order changing from $4\nu$ to $4(\nu+1/2)$ as the BL-like LLs cross the ML-like LLs~\cite{doi:10.1126/sciadv.aax6550, Zhou2023}. In the regions marked by \textbf{B', C}, the formation of  $C_3$ symmetric Dirac cones leads to the emergence of Landau level spaced by three (Fig.~\ref{fig:fig1}(f)).

\begin{figure*}[t]
	\includegraphics[width=\columnwidth]{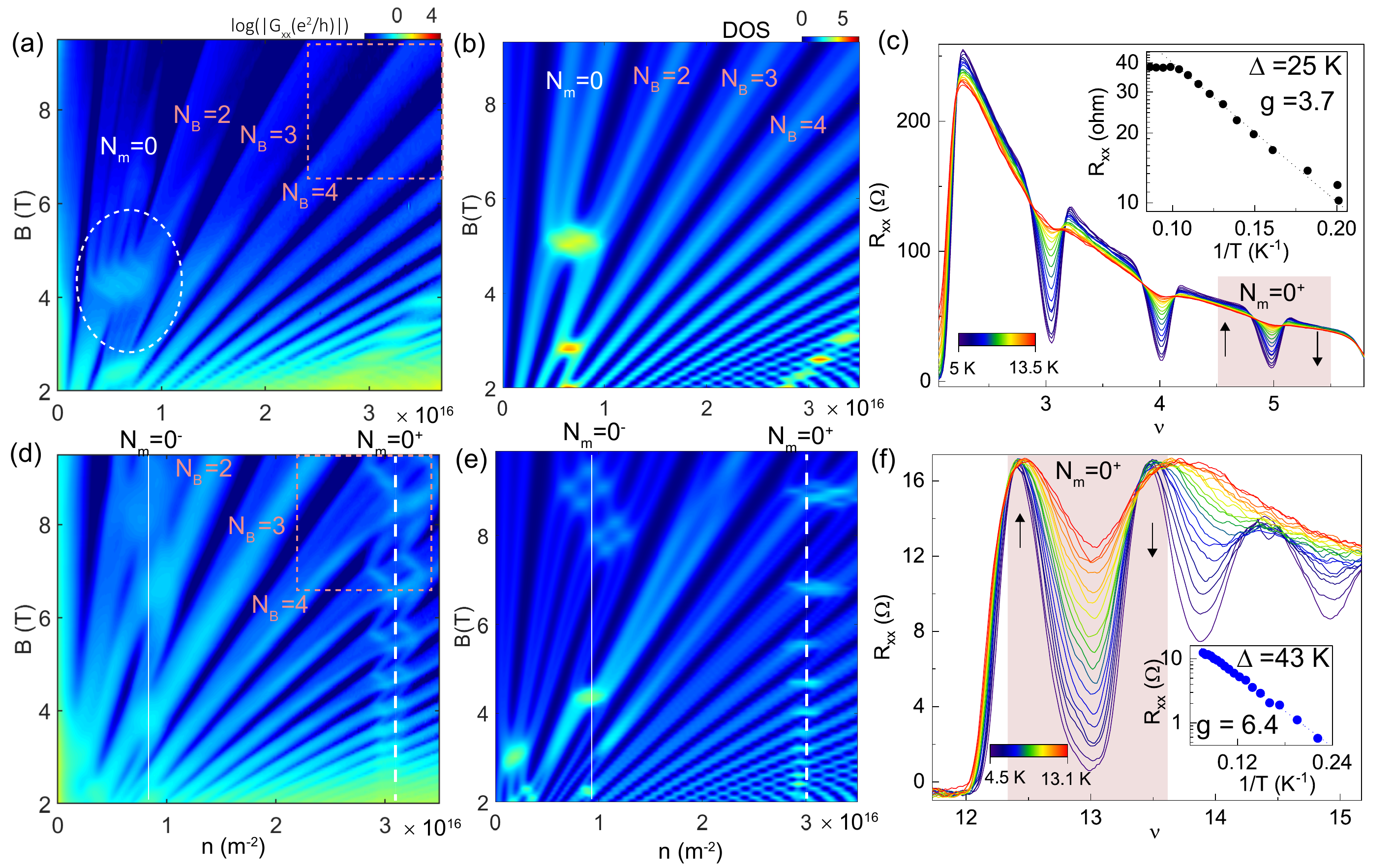}
	\caption{\textbf{Displacement field dependent fan diagrams}: Map of $G_{xx}$ as a function of $n$ and $B$ at (a) $D=0$ ~V/nm, and (d) $D = 0.75$ ~V/nm. The circle in (a) marks the crossing of $N_M=0$ LL with $N_B=2$ LL of the bilayer-like band. Simulated Landau level spectra for (b) $\Delta_1=0$ meV and (e) $\Delta_1=60$ meV. The vertical solid lines in (d) and in (e) indicate the positions of the spin-degenerate $\mathrm{N_M=0^-}$ ML-like LL. The solid dotted lines indicate the positions of the spin-degenerate $\mathrm{N_M=0^+}$ ML-like LL. Plot of $R_{xx}$ as a function of filling factor $\nu$ at different temperatures for (c) $D=0$ V/nm, (f) $D=0.75$ V/nm. The shaded region marks the $R_{xx}$ minima for $N_M=0^+$ LL. Insets show the Arrhenius fits to the minima of $R_{xx}$ as a function of $1/T$.}
	\label{fig:fig2}
\end{figure*}

Fig.~\ref{fig:fig2}(a) shows the Landau fan diagram at $D=0$ and $T=5$ K. The LLs of the bilayer sector emerge from $n=0$. The four LLs emerging from $n = 0.5 \times 10^{16}$~$ \mathrm {m^{-2}}$ are the spin and valley non-degenerate quartet of the zeroth LL of the ML band, labeled $N_{M}=0$~\cite{stepanov2016tunable, Datta2017}. The intersection of the $N_{M} = 0$ and the BL-like LL $N_{B} = 2$ near $B=4$~T gives rise to the pattern highlighted by the white dotted ellipse.

As noted before, $D$-field mixes the ML-like and BL-like bands, causing them to lose their pure ML-like and BL-like characters~\cite{PhysRevB.81.125304, PhysRevB.87.115422}. However, referencing the LLs at high $D$,  we retain the same notations at $D=0$ V/nm for consistency. In Fig.~\ref{fig:fig2}(d), we plot the Landau fan diagram measured at $D=0.75$ V/nm. We observe that the position of the zeroth LL of the ML-like band originating from the $K'$-valley, $N_{M}=0^{-}$ (marked by white solid line) changes little to $n \sim 0.7 \times 10^{16}$~$\mathrm {m^{-2}}$ as compared to its position at $D=0$. Conversely, the $N_{M}=0^{+}$ (zeroth LL of the ML-like band in the $K$ valley, marked by the dotted white line) moves to a significantly larger number density, $n \sim 3 \times 10^{16}$~$\mathrm {m^{-2}}$. To account for the observed behavior, we calculated the LL spectrum of ABA TLG using tight binding parameters based on Slonczewski-Weiss-McClure model parameters~\cite{PhysRevB.101.245411,PhysRev.109.272,MCCLURE1969425} with varying interlayer potential $\Delta_1$. For TLG, $\Delta_1 =-[(d_{\perp}/2\epsilon_{TLG})\times~D]e~$  \cite{PhysRevLett.132.096301, PhysRevLett.121.167601} leads to $\Delta_1 (\mathrm{eV}) = 0.084~D~(\mathrm{V/nm})$. Here, $d_{\perp}$=0.67 nm is the separation between the top and bottom layers of TLG, and $\epsilon_{TLG}$ is the dielectric constant of the TLG. The results are shown in Fig.~\ref{fig:fig2}(b) for $\Delta_1=0$~meV ($D=0$~V/nm) and Fig.~\ref{fig:fig2}(e) for $\Delta_1=60$~meV ($D \approx 0.75$~V/nm). They show that with increasing $D$, the conduction band of the monolayer sector (localized at the $K$-valley) drifts to higher energies, whereas the valence band of the monolayer sector (localized at the $K'$-valley) remains unchanged;   reproducing the observed experimental features. Note that in our calculations, LLs are spin non-degenerate with Land\'e g-factor of $2$.

To investigate the impact of LL crossings on the energetics of the system, we measured the Zeeman gap between the $N_M=0^+\uparrow$ and $N_M=0^+\downarrow$ LLs at different values of $D$. Fig.~\ref{fig:fig2}(c) and  Fig.~\ref{fig:fig2}(f) are the plots of longitudinal resistance $R_{xx}$ versus $\nu$ in the vicinity of $N_M=0^+$ LL measured at $D=0$~V/nm and $D = 0.75$~V/nm respectively for $B = 10~T$. The shaded rectangles mark the resistance minima for the Fermi energy lying between the $N_M=0^+\uparrow$ and $N_M=0^+\downarrow$ LLs.  An activated fit to the $R_{xx}$ minima gives an estimate of the gap $\Delta$ between the spin-up and spin-down  LLs: $R_{xx} = R_0\mathrm{exp}(-\Delta/2k_BT)$ with $\Delta =g^*\mu_BB$. We find $\Delta = 2.15$~meV and $3.71$~meV for $D= 0$ V/nm  and $0.75$~V/nm  respectively,  yielding $g^* \approx 4$ for $D = 0$~V/nm and $g^* \approx6.3$ for $D=0.75$~V/nm. Since activation measurements consistently underestimate the real gap ~\cite{PhysRevLett.132.046603}, these numbers represent a lower limit of $g^*(D)$. The observation of the $D$-dependence of spin splitting motivates defining an `effective' Land\'e $g$-factor 
$g^*\mu_BB=g\mu_BB+\mathcal{E}_{ex}^0(D) $
such that the  Zeeman energy $E_Z=\sigma g^*\mu_{B}B$ ($\sigma=\pm 1/2$ for the two opposite spins). Here, $\mu_{B}$ is Bohr magneton, $g=2$ is free electron $g$-factor, and $\mathcal{E}_{ex}^0(D)$ is the $D$-field dependent electron-electron interaction energy. In this picture, the increase in $g^*$ with $D$ reflects a commensurate increase in the electron-electron interaction energy.  With increasing $D$, the BL-like LLs and the $N_M=0^+$ LL approach each other. A strong electron-electron interaction will favor increased spacing between these LLs to minimize the repulsion between them. This, in turn, leads to an `effective' increase of the g-factor (see Supplementary Information, Section~\textbf{S4} for details).

Support for interaction-induced enhancement of LL spacing comes from examining the areas marked by red dotted rectangles in Fig.~\ref{fig:fig2}(a,d). In panel (a), the region is away from the crossing point of the BL-like and ML-like bands, and one observes that the BL-like LL are four-fold degenerate. In panel (d), by contrast, over the same range of magnetic fields but for $D=0.75$~V/nm, all the degeneracies of the BL-like bands are lifted, indicating an enhancement of the spacing between the BL-like LLs as they cross the ML-like LL (for details, see Supplementary section \textbf{S4}). This $D$-field induced lifting of the SU(4) symmetry of both BL-like and ML-like band was not seen previously in TLG~\cite{PhysRevB.86.155440} and provides us with the platform to tune the effective Land\'e g factor in the same system through a displacement field.

\begin{figure*}[t]
	\includegraphics[width=0.85\columnwidth]{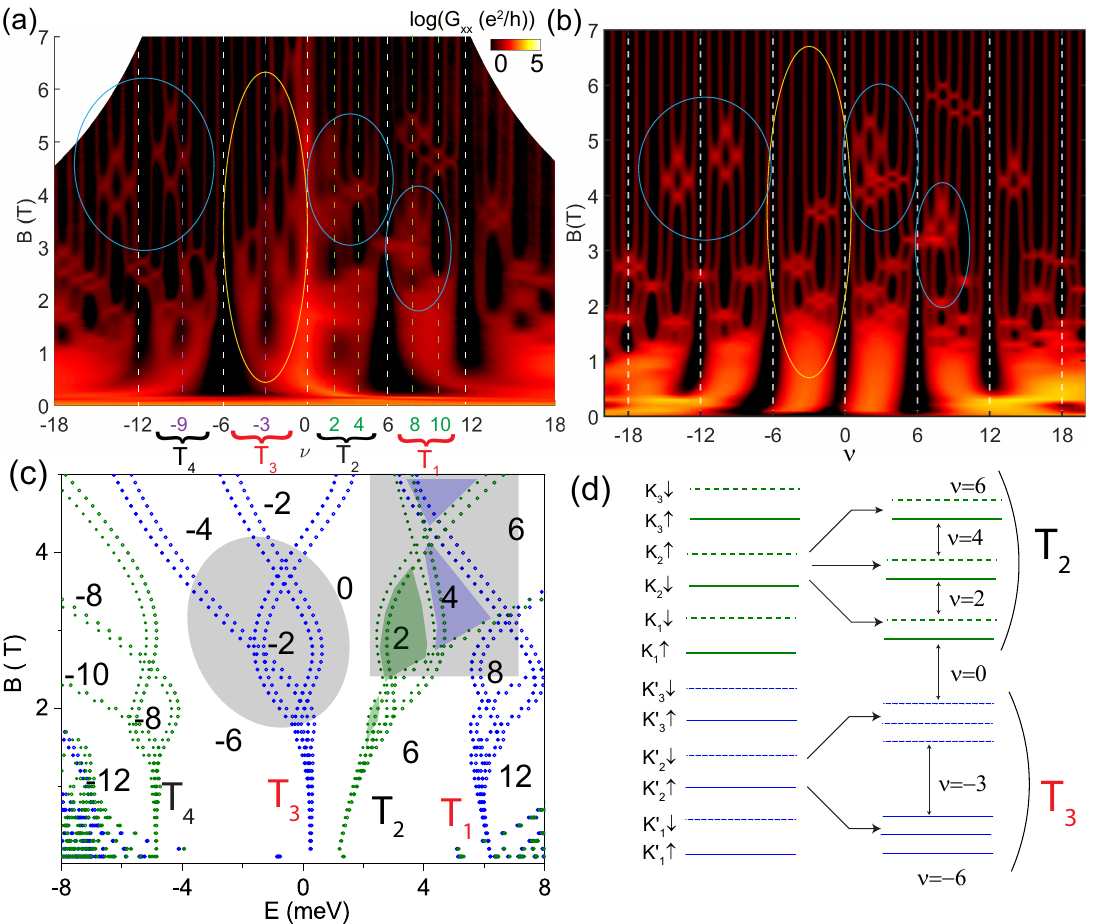}
	\caption{\textbf{Evolution of Landau levels near Dirac gullies}: (a) Measured contour map of $G_{xx}$ as a function of $\nu$ and $B$ measured at $D = 0.95$~V/nm. (b) Calculated contour map of the DOS as a function of $\nu$ and $B$ at $\Delta = 80$~meV. Here $D$ and $\Delta$ are related by $\Delta=\gamma$D with $\gamma=0.084$~e nm. The dark black regions correspond to vanishing DOS and hence conductance minima. The blue circles mark the regions of LL crossings in the experimental data and simulations. The yellow ellipse marks the region where simulated and experimental data diverge. (c) Simulated LL plot at $\Delta_1=80$ meV. $T_1$, $T_2$, $T_3$, $T_4$ are the gullies. The numbers in the plot are the filling factor $\nu$. (d) Schematic showing the inferred order of spin and gully degeneracy breaking in the two gullies $T_1$ (on the electron) and $T_2$ (on the hole side).}
	\label{fig:fig3}
\end{figure*}

Focusing on the BL-like band, Fig.~\ref{fig:fig3}(a) shows the contour plot of $G_{xx}$ as a function of magnetic field and filling factor $\nu$ at $D=0.95$ V/nm and $T=20$ mK. We observe multiple Landau level crossings that evolve with $B$. At low $B$, $G_{xx} = 0$ only at $\nu = \pm 6$ and $\pm 12$ (marked by white dashed lines), consistent with the emergence of six-fold degenerate states (three-fold Gully degenerate and two-fold spin degenerate). Increasing $B$ lifts the gully and spin degeneracies, leading to multiple phase transitions among these broken symmetry states. Theoretically predicted \cite{PhysRevB.87.115422, PhysRevB.101.245411}, these phases had escaped experimental determination.

The calculated density of states (DOS) is plotted in Fig.~\ref{fig:fig3}(b). The LL crossings observed in the experimental data are faithfully reproduced by our calculations  (regions marked by blue circles in Fig.~\ref{fig:fig3}(a) and (b)). From the simulated plots in Fig.~\ref{fig:fig3}(c), we identify the LLs between $\nu=12$ and $6$ emerging from $T_1$ gully, $\nu=6$ and $0$ emerging from $T_2$ gully, $\nu=0$ and $-6$ emerging from $T_3$ gully and those between $\nu=-6$ and $-12$ from $T_4$ gully.

As seen from Fig.~\ref{fig:fig3}(a), for positive energies, the first symmetry-broken LLs to be resolved are $\nu=2, 4, 6$ in the $T_2$ Gully and $\nu=8, 10, 12$ in the $T_1$ Gully;  these are marked by green dashed lines in Fig.~\ref{fig:fig3}(a). Conversely, for negative energies, the LLs initially appear at $\nu=-3$, and $-6$ in $T_3$ Gully and $\nu=-9$, and $-12$ in $T_4$ Gully (marked by purple dashed lines in Fig \ref{fig:fig3}(a)). The order of symmetry breaking of different gullies with the magnetic field is thus non-trivial and is strongly particle-hole asymmetric. While the experimentally observed LL crossings in the $\mathrm{T_1}$ and $\mathrm{T_2}$ Gullies conform to theoretical expectation (regions marked by blue ellipses in Fig.~\ref{fig:fig3}(a) and (b)), those in the $\mathrm{T_3}$ Gully deviate from theoretical expectations (regions marked by yellow ellipses in Fig.~\ref{fig:fig3}(a) and (b)). 

This electron-hole asymmetry in the order of symmetry breaking is unexpected. According to the Landau spectrum obtained from a single particle picture (calculated assuming $g=2$) presented in Fig.~\ref{fig:fig3}(c), the valley degeneracy is lifted first, resulting in QH states at $\nu= \pm6,~\pm12$. This is followed by the gully degeneracy lifting, which resolves additional filling factors at $\nu=\pm2,\pm4,\pm8,\pm 10$. This framework aligns well with our observations at positive energies where $G_{xx}$ minima occur at $\nu=2, 4, 8, 10$. However, the observation of $G_{xx}$ minima at $\nu=-3$ and $\nu = -9$ at negative energies suggests that for the holes, the spin degeneracy is lifted before the gully degeneracy (Fig.~\ref{fig:fig3}(d)). 

To address the observed discrepancy, we estimated the Land\'e g factor for the spin split LLs in both $T_2$ and $T_3$ gully (Supplementary Information, Section S6). We found the effective $g^*$ value for the $\mathrm{T_3}$ Gully is significantly enhanced to $6.31$ while for $\mathrm{T_2}$ Gully, it is $2.99$ which is closer to the bare value of $2$. This substantial increase in the value of $g^*$ for $T_3$ is a probable origin of the deviation of our experimental observations from the predictions of the non-interacting picture. 

\begin{figure}[t]
	\includegraphics[width=\columnwidth]{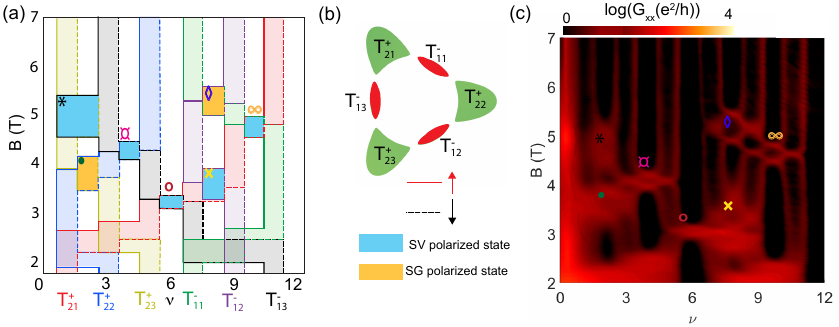}
	\caption{\textbf{Identification of filling sequence of QH states at high displacement fields:} (a) Evolution of LLs emerging from different gullies as a function of magnetic field and filling factor. The solid (dashed) line represents up spin LL (down spin LL). The cyan region hosts the spin and valley polarized states, and the yellow region hosts the spin and gully polarized states. Here, $T_{\alpha \beta}^{\gamma}$ represent the LL of the Gully with $\gamma$ as the valley index, $\alpha$ as the gully type, $\beta$ is the index of LL within the single Gully. (b) Calculated Fermi contour at $\Delta_1 =80$ meV.(c) Contour plot of $G_{xx}$ as a function of the magnetic field $B$ and filling factor $\nu$.} 
	
	\label{fig:fig4}
\end{figure}

Based on our experimental data and simulations, we present a phase diagram of the Gully region in the $\nu-B$ space in Fig.~\ref{fig:fig4}(a). $T_{\alpha\beta}^\gamma$ represents the Gully LL with $\alpha = 1,2$ the Gully index, $\beta = 1, 2, 3$ the component of the triplet and $\gamma = \pm$ the valley index (Fig.~\ref{fig:fig4}(b)). 
The solid lines (dashed lines) represent the spin-up (spin-down) LLs. Increasing $B$ at a fixed value of $\Delta_1 = 80$~meV leads to the braiding of the LLs, as shown in Fig.~\ref{fig:fig3}(c), causing multiple phase transitions among the LLs having different spin, valley, and Gully indices. Each symbol in Fig \ref{fig:fig4}(c) corresponds to the pattern shown in Fig \ref{fig:fig4}(a) and tracks the evolution of each LL with magnetic fields. The cyan-colored regions in Fig.~\ref{fig:fig4}(a) are the regions where the LLs are simultaneously spin and valley polarized. These are the regions where spin-split LLs from $T_1$ and $T_2$ gullies coexist. Similarly, the yellow areas host the spin-gully polarized state; here, spin-split LLs within the single Gully coexist. The overlap between the calculated spectrum in Fig.~\ref{fig:fig4}(a) and the experimentally measured one (Fig.~\ref{fig:fig4}(c)) is apparent from the one-to-one correspondence between the regions marked by different symbols in the two plots.

\section{Conclusions}

In conclusion, we have investigated the symmetry-broken quantum Hall states in both monolayer and bilayer sectors of Bernal-stacked trilayer graphene at high displacement fields. In the monolayer band, high displacement fields enhance the spin splitting gap for $N_M=0^+$ LLs. We attribute this enhancement to strong electron-electron interactions between ML-like and BL-like Landau levels, which additionally lead to complete degeneracy lifting of bilayer LLs near the crossing points.  We have also explored the symmetry-breaking sequence of LLs in the Dirac Gullies at high displacement fields and found a non-trivial gully polarized state with a strong electron-hole asymmetry. The displacement field dependence of the Zeeman splitting in trilayer graphene emphasizes the inadequacy of the often-used non-interacting model in studying quantum transport in multiband systems.

\section{AUTHOR INFORMATION}

S. K. and A. B. conceptualized the study, performed the measurements, and analyzed the data. U.G., A.S., and R.S. performed the theoretical analysis. K.W. and T.T. grew the hBN single crystals. S.K., A.B., U.G., A.S., and R.S. contributed to preparing the manuscript.

\section{Competing interests}

The authors declare no competing financial interest.

\begin{acknowledgement}
	A.B. acknowledges funding from U.S. Army DEVCOM Indo-Pacific (Project number: FA5209   22P0166). K.W. and T.T. acknowledge support from JSPS KAKENHI (Grant Numbers 19H05790, 20H00354, and 21H05233)
\end{acknowledgement}

\begin{suppinfo}
	
	Supporting information contains detailed discussions of  (a) device fabrication and characterization details and (b) data on other devices.
	
\end{suppinfo}

\clearpage

\clearpage

\section{Supplementary Information}

\renewcommand{\theequation}{S\arabic{equation}}
\renewcommand{\thesection}{S\arabic{section}}
\renewcommand{\thefigure}{S\arabic{figure}}
\renewcommand{\thetable}{S\arabic{table}}
\setcounter{table}{0}
\setcounter{figure}{0}
\setcounter{equation}{0}
\setcounter{section}{0}

\section{Device fabrication, schematics and characterization} 

\begin{figure*}[h]
	\includegraphics[width=0.8\columnwidth]{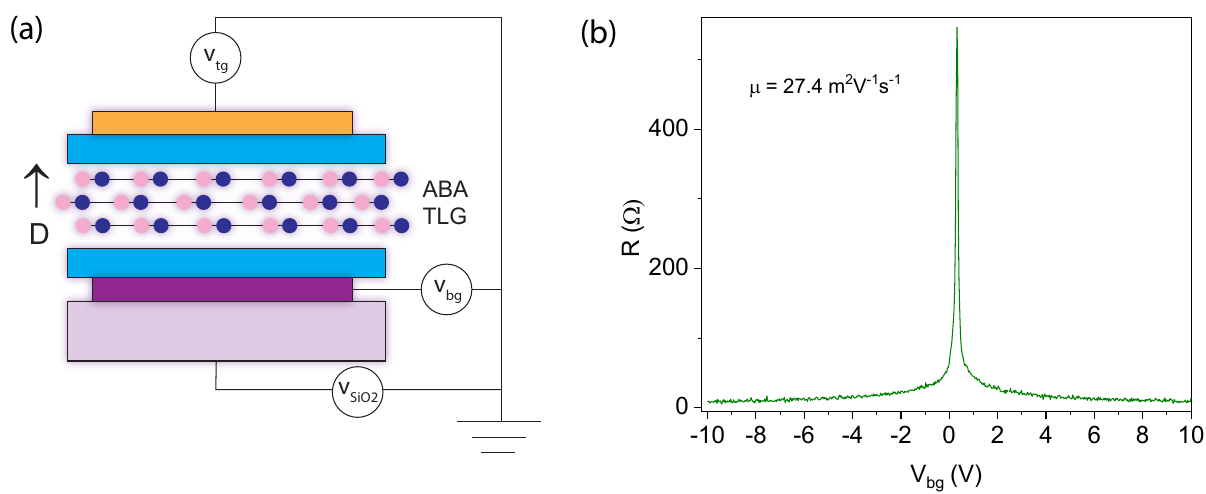}
	\caption{\textbf{Device schematic and characteristics:} (a) Schematic of ABA TLG encapsulated between two hBN flakes. (b) Plot of longitudinal resistance as a function of back-gate voltage measured at $B= 0$ T and $T = 0.02$ K. }
	\label{fig:figS1}
\end{figure*}

We mechanically exfoliated ABA-stacked trilayer graphene (TLG), hBN, and graphite flakes on  $300~$nm thick SiO$_2$/Si substrates. Flakes of ABA TLG were identified using optical contrast and Raman spectroscopy ~\cite{Cong2011, PhysRevLett.129.186802}.  Standard dry pickup and transfer techniques \cite{pizzocchero2016hot,doi:10.1126/science.1244358, Kaur2024} were used to fabricate the hBN/TLG/hBN/graphite heterostructure.  1-D contacts were defined using e-beam lithography and reactive ion etching using $\mathrm{CHF_3/O_2}$. We used Cr/Au ($5$~nm/$60$~nm) to define metallic contacts by thermal evaporation. The device was finally etched into a Hall bar shape. A device schematic with different gates is shown in Fig.~\ref{fig:figS1}(a). Graphite is used as the bottom gate, gold is the top gate, and the Si back gate is used to dope the graphene contacts to avoid the formation of a pn junction \cite{Kaur2024}.  Dual gate configuration is used to simultaneously tune the vertical displacement field $D=[(C_{bg}V_{bg}-C_{tg}V_{tg})/2\epsilon_0+D_{0}]$ and number density  $n=[(C_{bg}V_{bg}+C_{tg}V_{tg})/e+n_{0}]$ across the sample independently.  

A standard low-frequency lock-in detection technique is used to perform electrical transport measurements. Fig.~\ref{fig:figS1}(b) shows the resistance versus gate voltage response of the sample.  The impurity density of this device is $n_0 = 1.76 \times$ 10$^{10}$ cm$^{-2}$, and the  $\mu= 27.4~\mathrm{m^2V^{-1}s^{-1}}$ \cite{10.1063/1.3592338}. 
\section{Probing Lifshitz transitions in ABA TLG through transport studies.}

\begin{figure*}[h]
	\includegraphics[width=\columnwidth]{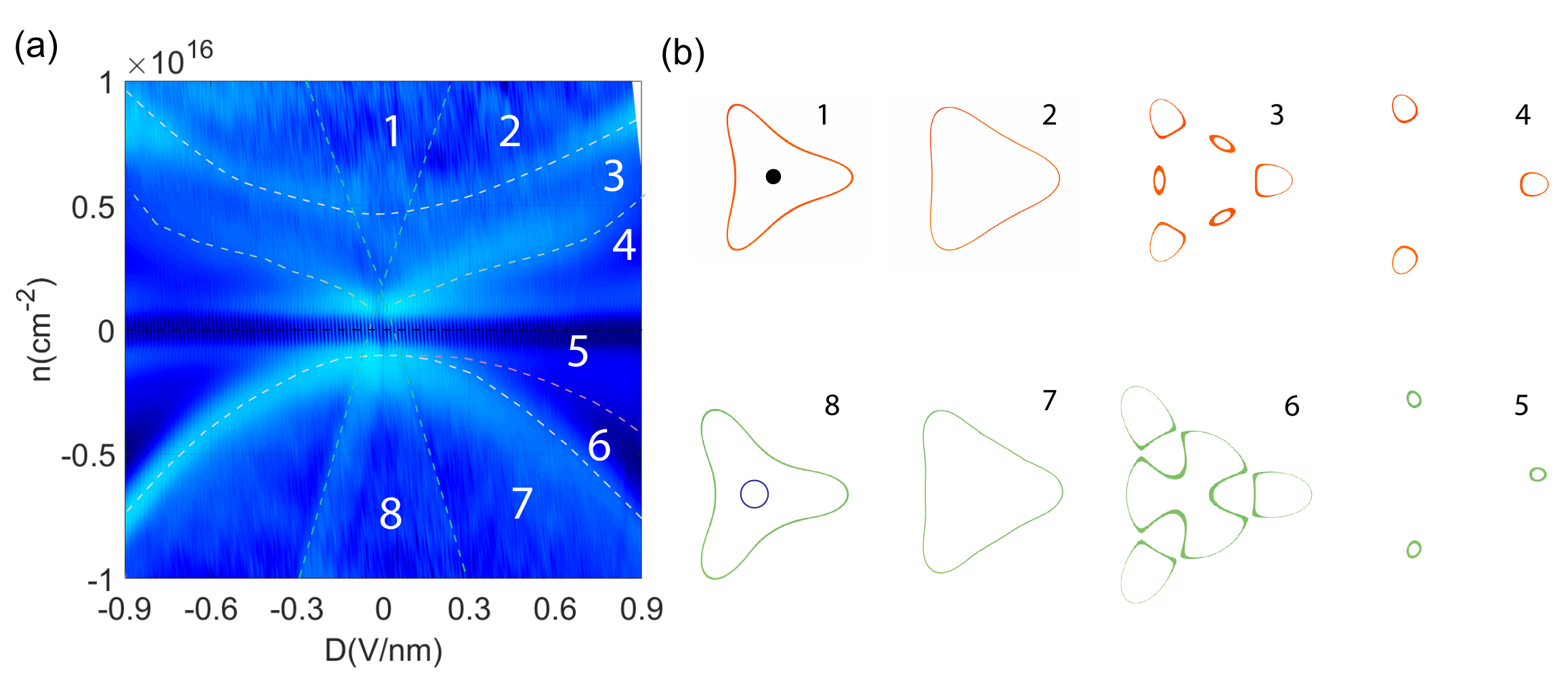}
	\caption{\textbf{Lifshitz transition in ABA TLG.} (a) Color plot of $dG_{xx}/dn$ as a function of number density and displacement field at $B=0$ T and $T=20$ mK. (b) Calculated Fermi contours of ABA trilayer graphene in different regions between Lifshitz transitions. The numbers indicated against each plot correlate to the labeled areas in (a). }
	\label{fig:figS4}
\end{figure*}

Fig.~\ref{fig:figS4}(a) shows the contour plot of the derivative of measured conductance $(dG_{xx}/dn)$ as a function of the displacement field and number density at $T=20$ mK and $B=0$ T. Various discontinuities in the $dG_{xx}/dn$ are seen; these are identified with the $D$-induced Lifshitz transitions in TLG.

Fig.~\ref{fig:figS4}(b) shows the calculated fermi surface contours in the different regions marked by dashed lines in Fig.~\ref{fig:figS4}(a). $1$ and $8$ mark the areas formed by monolayer (filled circle in $1$ and  open circle in $8$) and bilayer-like bands (orange contour in $1$ and green contour in $8$) at low displacement fields. Regions $2$ (electron side) and $7$ (hole side) host the single bilayer-like band at high $D$.  Region $3$ has two types of Gullies ($T_1$ and $T_2$) over the electron side at high $D$ and close to CNP. Areas $4$ and $5$ show the Gullies close to CNP. Region $6$ marks the formation of $T_3$ Gullies, whereas Gullies of $T_4$ merge together to form a single Fermi contour.

\section{Data from device B3S9}

\begin{figure*}[h]
	\includegraphics[width=\columnwidth]{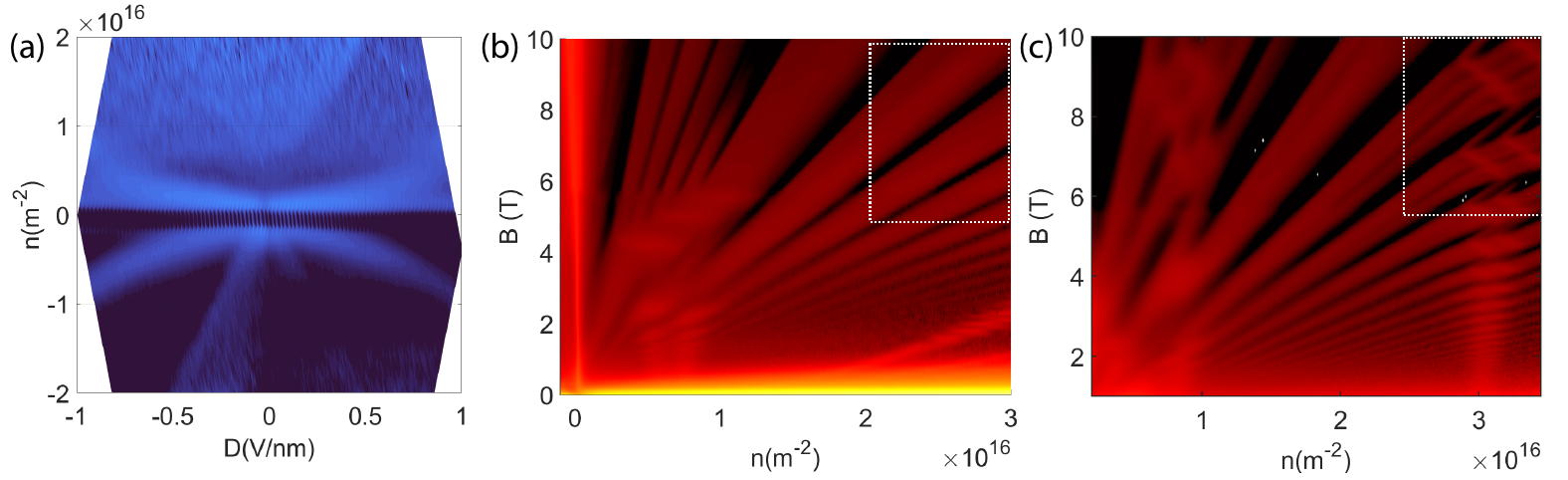}
	\caption{\textbf{Data from device, B3S9} (a) Color plot of d$G_{xx}/$d$n$ as  a function of number density and displacement field. Landau level fan diagram as a function of number density and magnetic field for (b) $D = 0$ V/nm, and  (c) $D = 0.75$~V/nm. }
	\label{fig:figS2}
\end{figure*}

Fig.~\ref{fig:figS2} shows the observations from another graphite-gated TLG device. Fig.~\ref{fig:figS2}(a) shows the color plot of d$G_{xx}/$d$n$ as a function of number density and displacement field. Various lifshitz transitions are visible as discontinuities in the plot. Fig.~\ref{fig:figS2}(b) and (c) are the Landau level fan diagrams measured at $D = 0$ V/nm and $0.75$ V/nm, respectively. In the dashed rectangular region in Fig.~\ref{fig:figS2}(c), the lifting of $4$-fold degeneracy in BLG bands is clearly visible.

\section{Effect of Landau level crossing on the ML-like and BL-like LLs}

\begin{figure*}[t]
	\includegraphics[width=0.85\columnwidth]{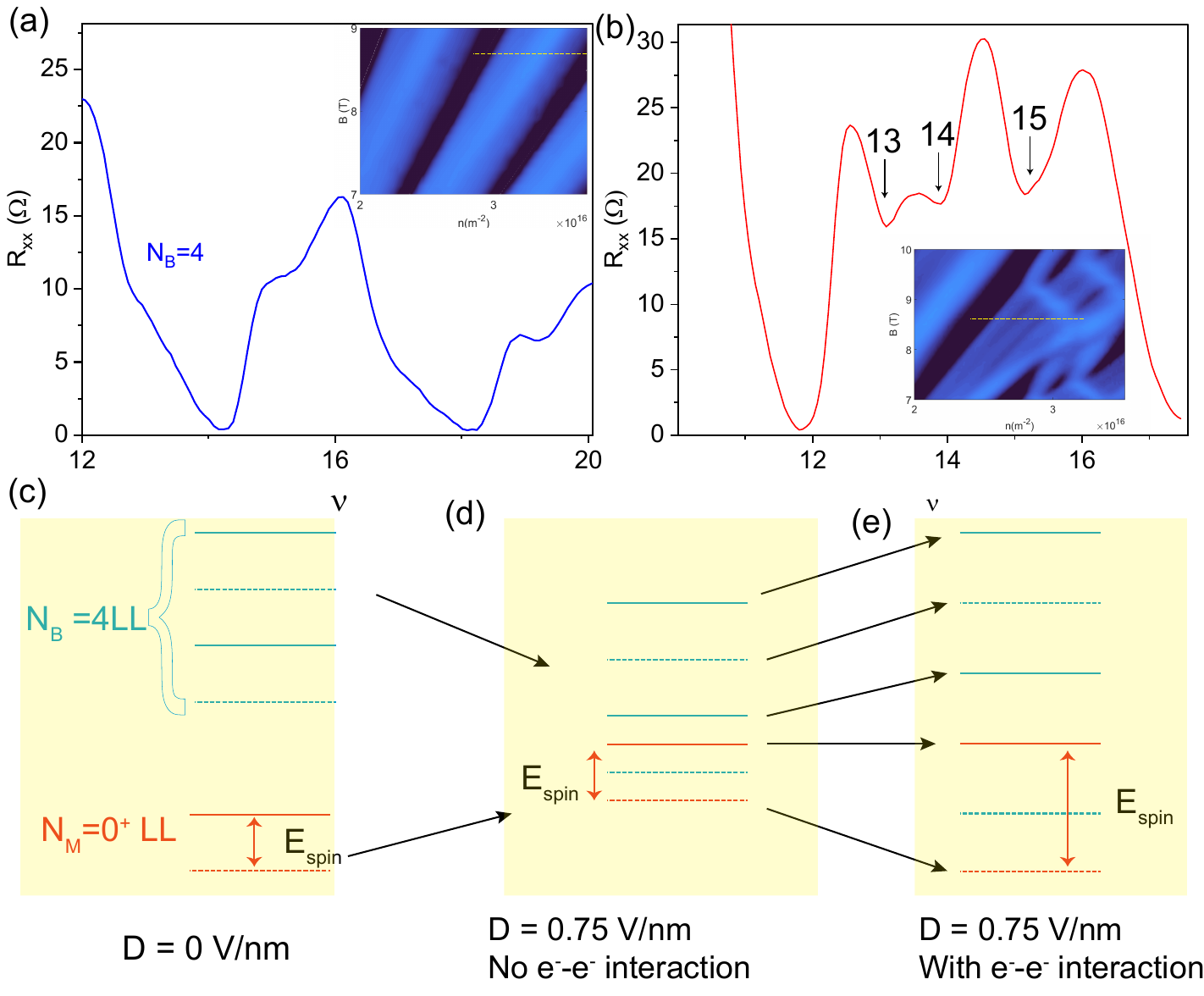}
	\caption{\textbf{Landau levels close to the crossing region of ML-like and BL-like LLs.} Plot of $R_{xx}$ as a function of filling factor $\nu$ measured at $B=8.7$ T and $T=5$ K for (a) $D=0$ V/nm (b) $D=0.75$ V/nm. Insets show the contour plot of $R_{xx}$ as a function of $B$ and $\nu$ close to the crossing region. (c) Schematic showing LLs of $N_B = 4$ and $N_M=0^+$ at $D=0$ V/nm. $\mathrm{E_{spin}}$ is the energy difference between up and down spin split Landau level of $N_M=0^+$. (d) Schematic showing LLs of $N_B = 4$ and $N_M=0^+$ at $D=0.75$ V/nm with no electron-electron interactions between ML and BL-like LLs. (e) Schematic showing LLs of $N_B = 4$ and $N_M=0^+$ at $D=0.75$ V/nm with strong electron-electron interactions between ML and BL-like LLs.}
	\label{fig:figS3}
\end{figure*}

Fig.~\ref{fig:figS3}(a) and (b) show plots of $R_{xx}$ as a function of filling factor $\nu$ for $D= 0$ V/nm and $D=0.75$ V/nm, respectively. These were measured at $T=5$ K and $B=8.7$ T. The arrows in Fig.~\ref{fig:figS3}(b) show the broken symmetry quantum Hall states of $N_B=4$ LL; these are absent in Fig.~\ref{fig:figS3}(a). The inset of Fig.~\ref{fig:figS3}(a) shows the contour plot of $R_{xx}$ in the vicinity of $N_B=4$ LL. Inset in Fig.~\ref{fig:figS3}(b) shows the contour plot of $R_{xx}$ in the vicinity of $N_B=4$ LL in the same range of $B$ and $n$ where it crosses the $N_M=0^+$ LL. The symmetry-broken states can be seen clearly in the inset close to the crossing region of $N_M=0^+$ and $N_B=4$ LL. 

To understand these observations, in Fig.~\ref{fig:figS3}(c--e), we schematically show the $N_B=4$ LL and $N_M=0^+$ LLs for three cases. Fig.~\ref{fig:figS3}(c) shows the case of $D=0$~V/nm, where two sets of LLs are far apart. We do not observe any symmetry breaking in this scenario. Fig.~\ref{fig:figS3}(d) shows the case of $D \neq 0$ V/nm with no interaction between $N_B=4$ and $N_M=0^+$ LLs. The $N_B=4$ LL and $N_M=0^+$ LL are now interleaved. Fig.~\ref{fig:figS5}(e) shows the case of $D \neq 0$ V/nm in the presence of significant electron-electron interactions. To minimize the Coulomb repulsion, the gap of LLs of $N_B=4$ and $N_M=0^+$ LL must be enhanced. This causes the increase in the spin gap of $N_M=0^+$ LL (shown in Fig.~2(e) and (f) of the main manuscript) and the observed symmetry-broken states of $N_B=4$ LL.

\section{Comparison of order of symmetry breaking at two different $D$.}

Fig.~\ref{fig:figS5} shows the contour plot of $G_{xx}$ at $D=0.95$ V/nm and $D=0.75$ V/nm measured at $T=20$ mK. The simulated DOS plots of LLs as a function of filling factor $\nu$ and magnetic field $B$ at $\Delta_1=80$ meV and $\Delta_1=63$ meV are also shown in Fig \ref{fig:figS5}. Table~\ref{table:tableS1} compares the values of $B$ at which the degeneracy between different Gullies break for $D=0.75$ V/nm labeled as $B_{gully}$ and $B_{spin}$  with the theoretical simulated values. In $T_1$, $T_3$ and $T_4$ Gullies the value of  $B_{gully}(D=0.75)~\mathrm{V/nm} < B_{gully}(D=0.95)~\mathrm{V/nm}$, in agreement with theoretical estimates. Conversely, for $T_2$, $B_{gully}(D=0.75)~\mathrm{V/nm} > B_{gully}(D=0.95)~\mathrm{V/nm}$, in contrast to theoretical expectations. We do not understand this discrepancy.

\begin{table}[h]
	\centering
	\resizebox{0.8\columnwidth}{!}{
		\begin{tabular}{| p{.2\textwidth} | p{.20\textwidth} | p{.2\textwidth} | p{.20\textwidth} |p{.20\textwidth} |} 
			\hline
			--  &  $B_{gully}$($D$=0.75) V/nm &  $B_{gully}$($D$=0.95 V/nm) & $B_{gully}$ ($\Delta_1=63$ meV) & $B_{gully}$($\Delta_1=80$ meV)\\
			\hline 
			
			$T_{1}$ & $1.70$ T & $2.21$ T &  $1.2$ T & $1.5$ T \\ 
			\hline 
			
			$T_{2}$ & $2$ T & $0.785$ T & $1.6$ T & $1.9$ T \\ 
			\hline
			$T_{3}$   &  $2.41$ T &  $2.45$ T & $1.6$ T & $1.9$ T \\
			\hline 
			$T_{4}$ & $1.46$ T & $1.71$ T &  $1.2$ T & $1.5$ T \\ 
			\hline 
			
	\end{tabular}}
	\caption{Values of magnetic field where Gully degeneracies lift for all Gullies.}
	\label{table:tableS1}
\end{table}

\section{Estimation of Land\'e \MakeLowercase{g} factor in $T_2$ and $T_3$ Gully at $D=0.75$ V/nm in another device.}

\begin{figure*}[t]
	\includegraphics[width=0.95\columnwidth]{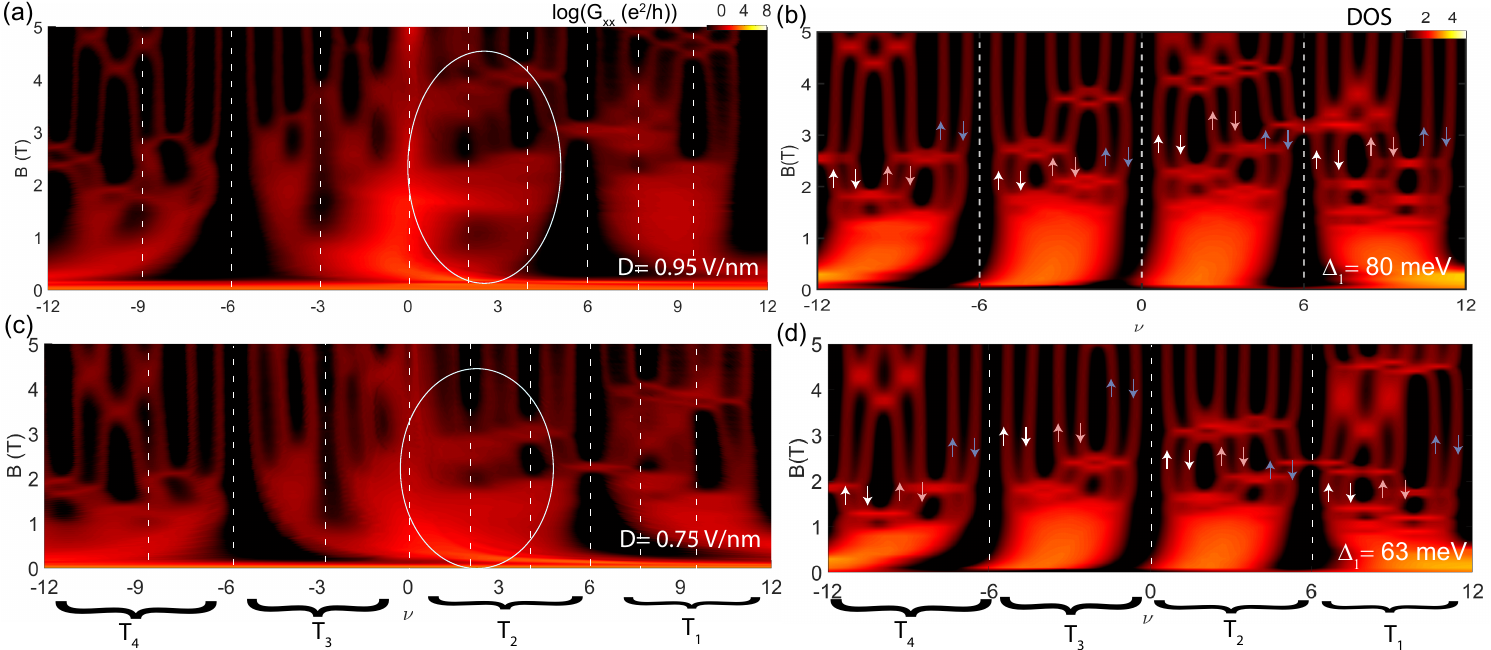}
	\caption{\textbf{Comparison of the symmetry breaking order of Gully LLs in different electric fields.} (a) Contour plot of $G_{xx}$ as a function of magnetic field and filling factor $\nu$ (a) for $D=0.95$ V/nm (c) for $D=0.75$ V/nm. The dashed lines mark the LLs appearing at the lowest magnetic fields. The circle in (a) marks the region with symmetry breaking at $\nu=2$ and indicates its absence in (c). Calculated DOS plot as a function of filling factor $\nu$ using single particle picture at (b) $\Delta_1=80$ meV and (d) $\Delta_1=63$ meV.}
	\label{fig:figS5}
\end{figure*}

\begin{figure*}[h]
	\includegraphics[width=\columnwidth]{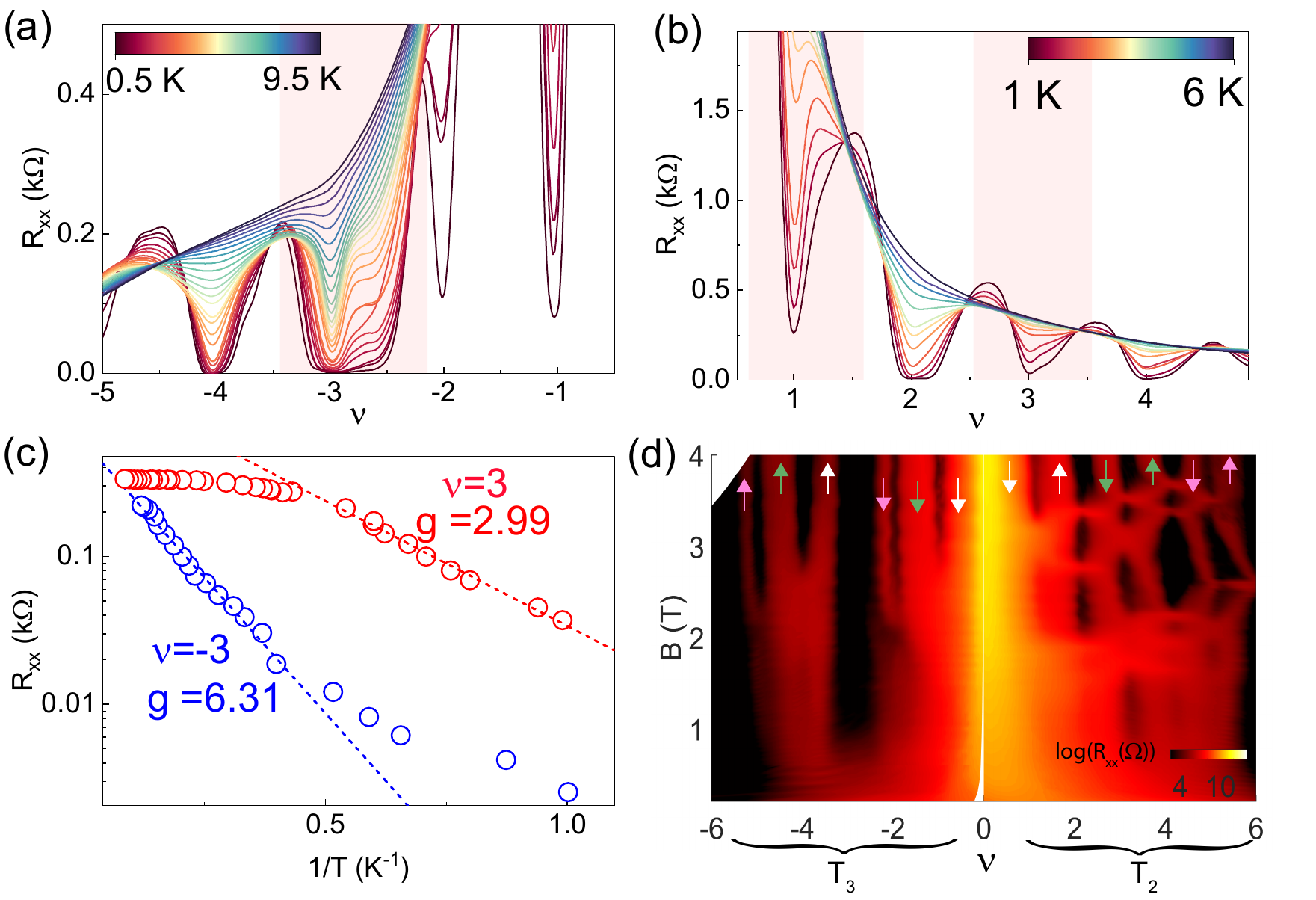}
	\caption{\textbf{Comparison of Land\'e g factor for $T_2$ and $T_3$ Gullies.} Plot of $R_{xx}$ as a function of filling factor $\nu$ for (a) $T_3$, (b) $T_2$ Gully measured at $D=0.75$ V/nm and $B=4$ T. The shaded region marks the resistance dip between spin-split LLs. (c) Estimation of Land\'e g factor for $T_2$ and $T_3$ Gully using Arrhenius fit to the data points. (d) Contour plot of $R_{xx}$ as a 
		function of filling factor and magnetic field at $D=0.75$ V/nm for another device. Here, the up and down arrow marks the spin up and spin down Landau levels, and each different color marks the LLs from 3 isolated off-center Dirac points in a Gully.}
	\label{fig:figS6}
\end{figure*}

To explain the observed violation of the order of symmetry breaking and absence of theoretically predicted landau level crossings for $T_3$ Gully (shown in Fig 3(a) of the main text and Fig \ref{fig:figS5}(a), (b)), we have estimated the Land\'e g factor for both positive $T_2$ and negative energies $T_3$. Fig \ref{fig:figS6}(a) and (b) shows the plot of $R_{xx}$ as a function of $\nu$ for $T_2$ and $T_3$ Gully at different temperatures. We have estimated the Land\'e g-factor from the Arrhenius fit to the value of $R_{xx}$ at the dip marked by shaded regions in Fig \ref{fig:figS6}(a) and (b). The plot is shown in Fig \ref{fig:figS6}(c). Here, the dashed lines are fit to the equation
\begin{equation}
	R_{xx}=R_0 \mathrm{exp}(-\frac{\Delta}{2k_BT})
\end{equation}
where,
\begin{equation}
	\Delta=g\mu_BB
\end{equation}

The estimated Land\'e g factor is $6.31$ for $T_3$ Gully and $2.99$ for $T_2$ Gully from the fits at filling factor $\nu= \pm3$. This significant increase of g-factor for $T_3$ over the bare value of $g$ in the spin-polarized state results in spin degeneracy breaking of the LLs in the Gullies followed by Gully degeneracy breaking. This further causes the violation of an order of symmetry breaking and the absence of multiple phase transitions between the LLs of gullies as expected from the non-interacting picture shown in Fig \ref{fig:figS5}(b) and (d). However, the estimated g-factor for $T_2$ Gully matches quite well with the bare g value, and observed crossings observed experimentally also match pretty well with the non-interacting picture. Fig \ref{fig:figS6}(d) shows the contour plot of $R_{xx}$ from another device at $D=0.75$ V/nm, and it matches pretty well with the data Fig \ref{fig:figS5}(c) of device 1. This further confirms the reproducibility of the data and the behavior of $T_2$ and $T_3$ Gully in different devices.

\clearpage

\bibliography{arxiv}
	
\end{document}